\documentclass[twocolumn,aps,showpacs,floatfix, prl]{revtex4-2}
\usepackage{graphicx}
\usepackage{amsthm,amsfonts,amssymb,verbatim}
\usepackage[usenames]{color}    %for colours
\usepackage{colortbl}
\usepackage{calrsfs}
\usepackage{hyperref} % for links in the bibliography
\usepackage{subfigure}
\usepackage{pgfplots}

\usepackage{mathtools}
\usepackage{physics}
\usepackage{xcolor}
\usepackage{adjustbox}
\usepackage{placeins}
\usepackage[T1]{fontenc}
\usepackage{lipsum}
\usepackage{csquotes}

\usepackage{fancyhdr}
\usepackage{cmap} 				% 
\usepackage{siunitx} % for formating tables

\newcommand{\bp}{\boldsymbol{p}}
\newcommand{\bq}{\boldsymbol{q}}

\newcommand{\br}{\boldsymbol{r}}
\newcommand{\bx}{\boldsymbol{x}}
\newcommand{\hbr}{{\hat{\bf r}}}

\newcommand{\balpha}{\boldsymbol{\alpha}}

\newcommand{\tf}{\widetilde{f}}
\newcommand{\tg}{\widetilde{g}}
\newcommand{\ctheta}{{\theta}}
\begin{document}
\title{
QED Corrections in Unstable Vacuum}

\author{V.~A.~Zaytsev}
\affiliation{Max-Planck-Institut f\"ur Kernphysik, Saupfercheckweg 1, 69117 Heidelberg, Germany}
\author{V.~A.~Yerokhin}
\affiliation{Max-Planck-Institut f\"ur Kernphysik, Saupfercheckweg 1, 69117 Heidelberg, Germany}
\author{C.~H.~Keitel}
\affiliation{Max-Planck-Institut f\"ur Kernphysik, Saupfercheckweg 1, 69117 Heidelberg, Germany}
\author{N.~S.~Oreshkina}
\email{natalia.oreshkina@mpi-hd.mpg.de}
\affiliation{Max-Planck-Institut f\"ur Kernphysik, Saupfercheckweg 1, 69117 Heidelberg, Germany}
\date{\today} % Leave empty to omit the date

\begin{abstract}

Self-energy and vacuum polarization effects in quantum electrodynamics (QED) are calculated for the supercritical Coulomb field, where Dirac energy levels become embedded in the negative-energy continuum. In this regime, the quantum vacuum becomes unstable, resulting in spontaneous electron-positron pair creation.
By calculating the imaginary part of the QED correction, we gain access to an unexplored channel of vacuum instability: radiative spontaneous pair creation.
Our results show that this radiative channel is greatly enhanced in the vicinity of the threshold of the supercritical regime, providing evidence for nonperturbative effects with respect to the fine-structure constant $\alpha$. 
We therefore conjecture that the total probability of spontaneous pair creation could differ significantly from the predictions of Dirac theory, especially near the supercritical threshold. 

\end{abstract}

\maketitle
%
% ============================================================================
%
% =========================
% \section{Introduction}
% =========================

\begin{figure}%
\centering
\includegraphics[width=0.45\textwidth]{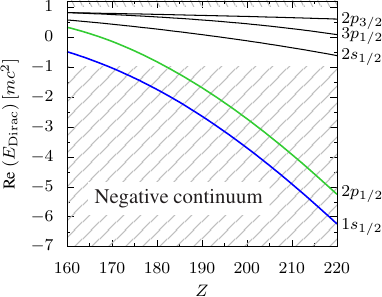}
\caption{
Real part of the Dirac  bound-state energies as a function of the nuclear charge number~$Z$. 
}
\label{fig_energy_cut}
\end{figure}

{\it Introduction. --- }
It has long been realized that for sufficiently large nuclear charges the energies of the lowest bound states of the Dirac equation may fall below the negative-continuum threshold, $E_D < -mc^2$, where $m$ is the electron mass, and $c$ is the speed of light.
This phenomenon emerges for nuclear charge numbers $Z > Z_{\rm cr}$$\,\approx\,$$173$ and is known as the {\it diving} of the Dirac bound states into the negative continuum (see  Fig.~\ref{fig_energy_cut}).
Once a bound state has undergone such diving, it is referred to as a {\it submerged} state.
The wave functions of these states are no longer square-integrable and their energies have an imaginary component, indicating that these originally bound states have become resonance states. 
This peculiar behavior is a manifestation of the instability of the quantum vacuum in the presence of a sufficiently strong field,  known as supercritical. 
The corresponding physical process is the creation of an electron-positron pair, in which the electron fills the vacancy in the submerged Dirac state and the positron is emitted into the continuum. 
This effect, known as spontaneous pair creation (SPC), was predicted long ago~\cite{Zeldovich, Greiner} but has not yet been observed experimentally.

Even after diving into the negative continuum, energies and widths of the Dirac states can nowadays be computed with high accuracy~\cite{Ackad_2007_1, Ackad_2007_2, Marsman_2011, Godunov_2017, Lv_PhysRevLett.121.183606, Maltsev2020}. 
The obtained widths are typically interpreted in terms of the probability of the SPC process~\cite{Fradkin}.
However, this interpretation assumes the smallness of quantum electrodynamics (QED) corrections or, equivalently, that the QED perturbative expansion in powers of the fine-structure constant 
$\alpha$ converges well.

QED is one of the most rigorously tested theories in modern physics. Owing to the smallness of the expansion parameter $\alpha \approx 1/137$, the QED perturbation series provides predictions of outstanding accuracy \cite{indelicato2019qed}. 
However, the convergence of  QED perturbation theory in the supercritical regime cannot be taken for granted. 
In fact, it was argued by Ritus~\cite{Ritus} and Narozhny~\cite{Narozhny} that for sufficiently strong constant crossed fields, the convergence of the QED  perturbation theory with respect to $\alpha$ breaks down, and QED becomes a fully nonperturbative and strongly coupled theory (see also reviews \cite{Akhmedov, Fedotov}).

Although the SPC process has never been observed so far, there is vast experimental effort directed to the creation and investigation of supercritical fields and unstable vacuum.
There are two main directions of these efforts. 
The first one involves high-power optical lasers \cite{Yoon:21, Mourou_ELI, Turcu_ELI, Apollon} and attempts to reach supercritical fields in laser-electron and electron-electron collisions \cite{Poder, Cole, PhysRevLett.122.190404, PhysRevD.104.L091903}. 
The theoretical investigations of such scenarios \cite{Podszus, Ilderton, DiPiazza} have not found confirmations of the Ritus-Narozhny conjecture so far. 
The second direction of experimental investigations of supercritical fields is connected to low-energetic collisions of two heavy nuclei. 
In a slow adiabatic collision, a quasimolecule can be formed by two colliding nuclei, with a enhanced nuclear Coulomb field into the supercritical regime. 
Such experimental studies are planned at a number of different facilities around the globe \cite{FAIR,HIAF,NICA}.

In this Letter, we explore the convergence of the QED perturbation theory in a supercritical Coulomb field. 
In the subcritical regime, QED corrections are typically suppressed by a factor of $\alpha$ relative to the Dirac energy. 
A nonperturbative character of QED in the supercritical regime would manifest itself as a breakdown of the expansion in $\alpha$, with QED corrections becoming comparable in magnitude to the Dirac energy. 
In order to test this conjecture, we calculate the first-order in $\alpha$ corrections to the Dirac energies in a supercritical Coulomb field, namely the electron self-energy and vacuum polarization. 
Moreover, knowledge of the imaginary part of these QED corrections gives us access to a new, radiative channel of spontaneous pair creation. 

% =========================
% \section{Complex scaling method}
% =========================
{\it Formalism. --- } There are two main problems in calculations of the QED corrections in a supercritical Coulomb field.
The first difficulty is that the electron-nucleus coupling parameter $Z\alpha$ is greater than 1, which requires a nonperturbative in $Z\alpha$  treatment of the electron propagators. 
The second difficulty is that the reference-state wave function is no longer square-integrable. This is the first calculation of radiative QED effects where both of these two problems must be solved simultaneously.

The starting point of our consideration is the stationary one-electron Dirac Hamiltonian:%
\begin{equation}
h_{D} = \balpha\cdot\bp
+ \beta m + V(r).
\label{eq_dirac_h}
\end{equation}
where $\balpha$ and $\beta$ are the Dirac matrices, $\bp$ is the momentum operator, and $V(r)$ a static nuclear-charge potential. 
We here consider the static spherically-symmetric nuclear field with $Z>Z_{\rm cr}$. 
Although such nuclei do not exist yet, this model approximates reasonaly well the two-center potential describing a slow collision of two nuclei with $Z_1 + Z_2 > Z_{\rm cr}$ at small distances, where the supercritical regime is supposed to be reached~\cite{artemyev:22}.

Solving the Dirac Hamiltonian for the bound states embedded into the negative-energy continuum is a nontrivial task since these states decay via spontaneous electron-positron pair creation and their wave functions are not square-integrable. 
A method of complex-scaling (CS) allows a systematic treatment of resonance states, see reviews~\cite{JUNKER1982207, HO19831, MOISEYEV1998212, LINDROTH2012247, SIMON1978, Riinhardt1982} and recent works~\cite{Zaytsev.PRA100.052504, Zaytsev.PRA107.032801}. 
The method is based on dilating the Dirac Hamiltonian into the complex plane of the radial variable $r$, e.g., by the uniform complex rotation
\begin{equation}
\br \rightarrow \br \, e^{i\ctheta},
\label{eq_r_ucr}
\end{equation}
leading to the following CS-transformed Hamiltonian
\begin{equation}
h_{D,\ctheta}
=
e^{-i\ctheta}\balpha\cdot\bp
+ \beta m + V(re^{i\ctheta}).
\label{eq_cs_hamiltonian}
\end{equation}

After the complex rotation, the Hamiltonian becomes a non-hermitian symmetric operator with complex eigenvalues. 
The energies of the true bound eigenstates of the rotated Hamiltonian remain unchanged. 
The energies of the continuum and the submerged states, however, acquire different imaginary parts and thus become separated from each other. 
The eigenvalues of the submerged states have the form $E = E_{\rm res} + i\Gamma/2$, where the real part $E_{\rm res}$ is the energy and the imaginary part $\Gamma$ is interpreted as the decay rate due to the spontaneous positron--bound-electron pair creation.
The corresponding calculations were carried out in Refs.~\cite{Ackad_2007_1, Ackad_2007_2, Marsman_2011}.

The crucial advantage of the CS method is that after the complex rotation the wave functions of the submerged states become square-integrable.
This opens a way not only for solving the Dirac equation but also for generalizing this method to calculations of the radiative QED corrections, which is accomplished in the present Letter. 

There are two QED corrections of the first order in $\alpha$, the electron self-energy and vacuum polarization, each of which will be addressed here.

% =========================
% \section{Self-energy}
% =========================

{\it Self-energy. --- } 
The self-energy correction of the state $a$ with energy $\varepsilon_a$ and wave-function $\psi_a$ is expressed as 
\begin{align}
\Delta E_{\rm SE} &= 
2i\alpha \int_{C_F} d\omega \int d\br_1 d\br_2 
\psi_a^\dagger(\br_1) 
\alpha_\mu G(\varepsilon_a - \omega,\br_1, \br_2) \notag \\ &\times \alpha_\nu 
\psi_a(\br_2) D^{\mu\nu}(\omega,\br_{12})
-
\delta m
\int d \br \psi^{\dag}_a(\br) \beta \psi_a(\br),
\label{eq_se}
\end{align}
where $\alpha_\mu = (1,\balpha)$, %\deleted[id=no]{$\balpha$ are the Dirac matrices,} 
$\delta m$ is the mass counterterm, $G$ is the Dirac-Coulomb Green's function, and $D^{\mu\nu}(\omega,\br_{12})$ denotes the photon propagator with $\br_{12} = \br_1 - \br_2$, and $C_F$ denotes the Feynman integration contour.

\begin{figure*}%[h!]
\includegraphics[width=\textwidth]{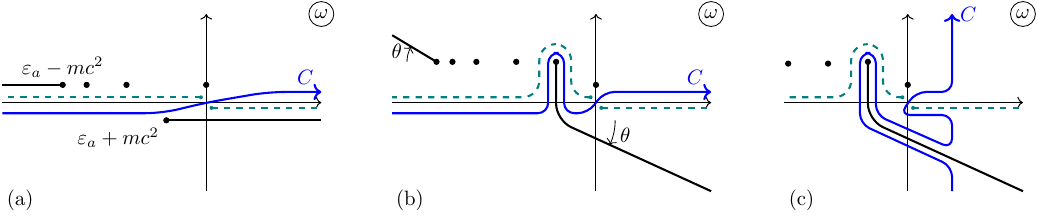}
\caption{The analytical structure of the integrand of the self-energy operator in the complex $\omega$ plane, for $a = 1s$ reference state, in the supercritical regime $\varepsilon_a < -mc^2$: a) before the complex rotation, b) after the complex rotation (with the reference-state energy $\varepsilon_a$ aquiring a positive imaginary part), c) after the deformation of the integration contour for the numerical integration.
The dashed turquoise line represent the branch cuts of the photon propagator. 
Dots and solid black line show the poles and branch cuts of the electron propagator, respectively. 
Solid blue line denotes the contour of $\omega$ integration. 
}
\label{fig_se_integCountour}
\end{figure*}

Eq.~(\ref{eq_se}) is only a formal expression since it contains ultraviolet divergences that need to be covariantly regularized and eliminated. 
This is achieved by isolating two first terms of the expansion of the Dirac Green function in powers of interaction with the binding potential~\cite{Yerokhin_PRA60_800, Oreshkina_PRA101_032511_2020}: 
\begin{equation}
    \Delta E_{\rm SE} = \Delta E^{(0)} + \Delta E^{(1)} + \Delta E^{(2+)},
\label{eq_se_012}\end{equation}
The first two terms, referred to as the zero- $\Delta E^{(0)}$ and one-potential $\Delta E^{(1)}$ terms, are calculated in momentum space within the dimentional regularization, whereas the remaining contribution, referred to as the many-potential term $\Delta E^{(2+)}$, is calculated in coordinate space with the partial-wave expansion of the Dirac Green function. 

When the energy of the reference state enters in the supercritical regime,  $\varepsilon_a < -mc^2$, the reference-state wave function ceases to be square-integrable and Eq.~(\ref{eq_se_012}) cannot be calculated by techniques developed in previous investigations. 
We overcome these difficulties by applying the complex rotation of the radial variable (\ref{eq_r_ucr}) and the corresponding rotation of the momentum, $p \rightarrow pe^{-i\ctheta}$. 
The complex-rotated Dirac Green function is obtained by solving the CS Dirac equation~\eqref{eq_cs_hamiltonian}.

One of the major difficulties in numerical evaluation of the self-energy correction comes from the fact that the complex rotation (\ref{eq_r_ucr}) greatly complicates the analytical structure of the integrand in Eq.~(\ref{eq_se}) as a function of the complex $\omega$, which entails the necessity of using elaborated integration contours in the complex $\omega$ plane. 
Fig.~\ref{fig_se_integCountour} shows the poles and branch cuts of the Dirac Green function, branch cuts of the photon propagator, and the integration contours used for the numerical evaluation of the self-energy correction in the supercritical regime.

We now present expressions for the individual terms in Eq.~(\ref{eq_se_012}) after the CS transformation. 
The zero-potential and one-potential terms are given by
\begin{align}
\Delta E^{(0)} &= 
e^{-3i\ctheta}
\int \frac{d\bp}{(2\pi)^3} 
\bar{\psi}_a(\bp e^{-i\ctheta}) \Sigma_R^{(0)}(\varepsilon_a;\bp e^{-i\ctheta}) %\notag \\ &\times 
\psi_a(\bp e^{-i\ctheta})
\label{eq_se_zero}
\end{align}
and
\begin{align}
\Delta E^{(1)} &\, = 
e^{-6i\ctheta}
\int \frac{d\bp}{(2\pi)^3} \int \frac{d\bp'}{(2\pi)^3}  
\bar{\psi}_a(\bp'e^{-i\ctheta}) \, V(\bq e^{-i\ctheta})
\notag \\ &\times
\Gamma_R^{0}(\varepsilon_a;\bp'e^{-i\ctheta},\bp e^{-i\ctheta})\, 
\psi_a(\bp e^{-i\ctheta}),
\label{eq_se_one}
\end{align}
where $\bar{\psi}_a = \psi^{\dag}_a\gamma^0$, $\bq = \bp'-\bp$ and
\begin{equation}
\psi_a(\bp e^{-i\ctheta}) = 
e^{3i\ctheta}
\int d\br e^{-i\bp\cdot\br} \psi_a(\br e^{i\ctheta}).
\label{eq_fourier}
\end{equation}
The explicit expressions for the renormalized free operators $\Sigma^{(0)}_R$ and $\Gamma_R^{0}$ can be found in Refs.~\cite{Yerokhin_PRA60_800, Oreshkina_PRA101_032511_2020}. 
Details are presented in the Supplementary Materials. 
The many-potential term is expressed as
\begin{align}
\Delta E^{(2+)} &= 2i\alpha e^{2i\ctheta} \int_{C_F} d\omega \int d\br_1 \int d\br_2 
\psi_a^\dagger(\br_1 e^{i\ctheta}) \alpha_\mu 
\notag \\ &\times 
G^{2+}(\varepsilon_a - \omega, \br_1 e^{i\ctheta}, \br_2 e^{i\ctheta})\alpha_\nu 
\psi_a(\br_2 e^{i\ctheta})
\notag \\ &\times
D^{\mu\nu}(\omega,\br_{12} e^{i\ctheta}),
\label{eq_many_pot}
\end{align}
where $G^{(2+)}(\omega) \equiv G(\omega) - G^{(0)}(\omega) - G^{(1)}(\omega)$ is the Dirac Green function containing two or more interactions with the binding potential. 
Details about the numerical evaluation of these terms and further formulas are given in Supplementary Materials.

% =========================
% \section{Vacuum polarization}
% =========================

{\it Vacuum polarization. --- }
The vacuum polarization (VP) potential is usually divided into the Uehling part, containing one interaction with the binding potential inside the lepton loop, and the Wichmann-Kroll part, containing three or more interactions with the binding potential inside the loop. 
For supercritical fields, both these terms are of the same order and need to be accounted for on the same footing. 

In the supercritical regime, $\varepsilon_a < -mc^2$, we need to perform the complex rotation in order to make the reference-state wave function squarely integrable. 
The Uehling potential then becomes
\begin{align}
\nonumber
&re^{i\ctheta}V_{\rm Ue}(re^{i\ctheta}) 
 = 
-\alpha(\alpha Z)\frac{2e^{2i\ctheta}}{3}
\int_0^\infty r' \rho_{\rm nuc}\left(r'e^{i\theta}\right) dr'
\\
& \times 
\int_1^\infty dt \left(1 + \frac{1}{2t^2}\right)\frac{\sqrt{t^2 - 1}}{t^3}
\left[
e^{-2 e^{i\ctheta} \left\vert r - r'\right\vert t }
-
e^{-2 e^{i\ctheta} \left( r + r'\right) t }
\right]\,,
\end{align}
where $\rho_{\rm nuc}(r)$ is the nuclear-charge density.
This expression contains only analytical functions in the integrand and thus can be calculated without problems. 

The Wichmann-Kroll contribution contains the Dirac Green function and is much more difficult to compute. 
After the complex rotation, we write the corresponding potential as
\begin{eqnarray}
V_{\rm WK}(\br e^{i\theta}) &= & \frac{\alpha}{2\pi i} 
\int d\br'  \frac{1}{|\br - \br'|} \int_{C_F} d\omega\, \\ &\times& {\rm Tr} 
\int e^{i\theta} d\bx G^{(0)}(\omega, \br' e^{i\theta}, \bx e^{i\theta}) V(x e^{i\theta})
\nonumber \\ & \times &
%V(\bx e^{i\theta}) 
\left[G(\omega, \bx e^{i\theta}, \br' e^{i\theta})
- G^{(0)}(\omega, \bx e^{i\theta}, \br' e^{i\theta})\right]. \nonumber
\end{eqnarray}
In order to evaluate the Wichmann-Kroll contribution, we make the Wick rotation of the $\omega$ integration contour, $\omega \to i\omega$ and evaluate separately the pole contributions due to the most deeply bound states. 

% =========================
% \section{Results}
% =========================
{\it Results. ---}
We now present our numerical results for the QED corrections to the energies of the two lowest-lying states in the supercritical regime.
We compute both the real part of the energies and their imaginary parts, which gives us access to the decay probability of these states.

\begin{figure*}%[h!]
\includegraphics[width=\textwidth]{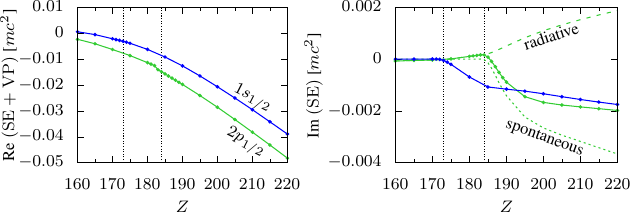}
\caption{
QED correction to the $1s$ and $2p_{1/2}$ energy levels as a function of the nuclear charge number $Z$, real part (left) and imaginary part (right).
}
\label{fig_se_vp}
\end{figure*}

The left panel of Fig.~\ref{fig_se_vp} presents our results obtained for the real part of the QED correction to the $1s_{1/2}$ and $2p_{1/2}$ energies, given by the sum of the self-energy and vacuum-polarization contributions. 
We observe that the QED correction is negative and therefore speeds up the diving of $1s_{1/2}$ and $2p_{1/2}$ levels into the negative continuum.

Of particular importance is the imaginary part of the QED correction, which is induced only by the self-energy contribution. 
Before diving, the imaginary part of the $1s_{1/2}$ self-energy is zero (since the ground Dirac state cannot decay), whereas for the $2p_{1/2}$ state it is nonzero and corresponds to the  probability of the radiative decay into the $1s_{1/2}$ state.  
After the diving, the $1s_{1/2}$-state self-energy acquires the imaginary part, which corresponds to the probability of a new process never considered so far in the literature, namely, the spontaneous pair creation accompanied by emission of a photon. 
We name this process {\it radiative spontaneous pair creation} (RSPC). 
One may view it as the process where an electron from the negative continuum with energy higher than the submerged state decays into it, emitting a photon and creating a positron.

Our numerical results for the imaginary parts of the $1s_{1/2}$ and $2p_{1/2}$ self-energy corrections are depicted in the right panel of Fig.~\ref{fig_se_vp}. 
We observe that for the $1s_{1/2}$ state, the imaginary part is zero before diving ($Z<Z_{\rm cr}\approx 173$), whereas in the supercritical regime, it monotonically increases with the nuclear charge, corresponding to the probability of the RSPC process.

The behavior of the imaginary part of the $2p_{1/2}$ self-energy correction is more complicated than for the $1s_{1/2}$ state.
Before the diving of the $2p_{1/2}$ state, which occurs at $Z\approx 184$, the imaginary part is nonzero but very small, signifying strong suppression of the $2p_{1/2}$-$1s_{1/2}$ decay. 
This suppression in the vicinity of the diving point is interesting because usually the decay probability grows with the increase of the transition energy and these two states are separated by almost 500~keV at this point. 
The suppression is explained by the fact that the radial parts of the $1s_{1/2}$ and $2p_{1/2}$ wave functions almost coincide for such large $Z$, which results in very small radial integrals for the one-photon E1 transition matrix elements.

After diving of the $2p_{1/2}$ state, there are two different decay channels. 
The first is the RSPC process, which proceeds similarly to that for the $1s_{1/2}$ state. 
Numerical results for this channel are depicted by the dotted green line in the right panel of Fig.~\ref{fig_se_vp} and labeled as ``spontaneous''. 
The second channel is the process in which the spontaneous pair creation occurs at the $1s_{1/2}$ state (assuming that it is still vacant) and at the same time the $2p_{1/2}$ electron decays into the $1s_{1/2}$ vacancy. 
This channel corresponds to the single-quantum electron-positron annihilation process and can be named the radiative decay to the negative continuum. 
Numerical results for this channel are depicted by the dashed green line in the right panel of Fig.~\ref{fig_se_vp} and labeled as ``radiative''. 
The radiative channel is of course absent in the scenario where the $1s_{1/2}$ state is already filled.
It is interesting that the radiative channel is present also in the intermediate regime when the $1s_{1/2}$ state is already submerged (but still vacant) and the $2p_{1/2}$ state has not yet dived. 

Now we turn to testing the Ritus-Narozhny conjecture about the possibility of nonperturbative in $\alpha$ effects  in the supercritical regime. 
To this end, we introduce the enhancement factor ${\cal R}$ as the normalized ratio of the probabilities of the radiative and non-radiative SPC processes, which is proportional to the ratio of the imaginary parts of the self-energy correction  ${\rm Im} \, E_{\rm SE}$ and the Dirac energy ${\rm Im} \, E_{\rm D}$,
\begin{align}\label{eq:ratio}
{\cal R} = \frac1{\alpha}\,\frac{{\rm Im} \,E_{\rm SE}}{{\rm Im}\, E_{\rm D}}\,.
\end{align}
Taking into account that the radiative channel is suppressed by the fine structure constant $\alpha \approx 1/137$ with respect to the nonradiative SPC channel, the enhancement factor ${\cal R}$ is expected to be of the order of one in the perturbative regime. 

\begin{table}
\begin{ruledtabular}
\begin{tabular}{cc}
%\hline\hline
$Z$ & ${\cal R}_{1s}$
\\ \hline
175 &  65.3\\
180 &  24.5\\
185 &  13.6\\
190 &   7.7\\
195 &   5.1\\
200 &   3.9\\
%\hline\hline
\end{tabular}
\end{ruledtabular}
\caption{The enhancement factor of the probabilities of the radiative and non-radiative SPC processes ${\cal R}$ defined by Eq.~(\ref{eq:ratio}) for the $1s$ state as a function of the nuclear charge number $Z$.}
\label{tab_rel}
\end{table}

In Table~\ref{tab_rel} we present the ratio ${\cal R}$ as a function of $Z$ in the supercritical regime for the $1s_{1/2}$ state. 
We observe that the enhancement factor of the radiative SPC process in the supercritical region is much larger than 1. 
Moreover, it rapidly increases when $Z$ approaches the diving point in the supercritical regime. 
This is an evidence that the nonperturbative regime with respect to $\alpha$ could emerge in the vicinity of the diving point. 

% =========================
%{\it Conclusions.}
% =========================
{\it Conclusions. ---}
We performed calculations of the complete set of QED effects of first order in $\alpha$ in the supercritical Coulomb field, in which the reference-state Dirac energy level is embedded in the negative-energy continuum. 
We demonstrated that the real part of the QED effects is negative and monotonically increases in magnitude with increase of the nuclear charge. 
This confirms the statement expressed in previous studies~\cite{Gyulassy1, MULLERNEHLER1994101, Malyshev_2022} that QED effects do not prevent Dirac levels from diving into the negative continuum. 

By calculating the imaginary part of the QED corrections, we obtained the probability of the radiative spontaneous pair creation, a process not yet addressed in the literature. 
Furthermore, an additional channel of the radiative decay of the $2p_{1/2}$ state to the negative continuum was identified. 

We found that the channel of the one-photon radiative spontaneous pair creation is greatly enhanced in the vicinity of the diving point, as compared to the nonradiative spontaneous pair creation. 
This evidence leads us to a conjecture that the nonperturbative regime with respect to the fine-structure constant $\alpha$ is realized in this region. 
It seems plausible that the situation near the diving point might be analogous to the well-known infrared catastrophe~\cite{Bloch_Nordsieck_1937, Yennie_1961, Jauch_Rohrlich_1976}, where the probability of emission of arbitrary numbers of soft photons is comparable and an expansion in the number of emitted photons does not converge. 
In our case, the consequence of the nonperturbative regime will be that the total probability of the spontaneous pair creation differs significantly from the predictions of the Dirac theory, especially near the supercritical threshold. 

% ========================

\bibliography{refs}

\widetext

\appendix
\newpage
\section{Complex-Scaled free-electron Dirac Green function}
\label{free_green}
The complex-scaled free-electron Dirac Green function can be expressed through the solutions being regular at the origin and in the infinite as follows
\begin{equation}
G^{(0)}_{\theta}(E,\br_1,\br_2) = -\sum_{\kappa m}
\left[ 
\Phi_{E\kappa m, \theta}^{\infty}(\br_1) \Phi_{E\kappa m, \theta}^{0\dagger}(\br_2)\theta(r_1 - r_2)
+
\Phi_{E\kappa m, \theta}^{0}(\br_1) \Phi_{E\kappa m, \theta}^{\infty\dagger}(\br_2)\theta(r_2 - r_1)
\right],
\end{equation}
where $\Phi_{E\kappa m, \theta}^{0}$ and $\Phi_{E\kappa m, \theta}^{\infty}$ has the following form 
\begin{equation}
\Phi_{E\kappa m, \theta}(\br) = \frac{e^{-i\theta}}{r}
\left(\begin{aligned}
G_{E\kappa}(re^{i\theta}) \Omega_{\kappa m}(\hbr)
\\
iF_{E\kappa}(re^{i\theta}) \Omega_{-\kappa m}(\hbr)
\end{aligned}\right)
\end{equation}
and correspond to the solutions of the complex-scaled Dirac equation
\begin{equation}
\begin{pmatrix}
\frac{d}{dr} + \frac{\kappa}{r} && -e^{i \theta} (1 + E)
\\
-e^{i \theta} (1 - E) && \frac{d}{dr} - \frac{\kappa}{r}
\end{pmatrix}
\begin{pmatrix} G_{E\kappa,\theta} \\ F_{E\kappa,\theta} \end{pmatrix} = 0
\end{equation}
being regular at the origin and infinity, respectively.
The explicit form of these solutions are
\begin{equation}
\begin{pmatrix} G^0_{E\kappa,\theta} \\ F^0_{E\kappa,\theta} \end{pmatrix} =
r e^{i\theta} 
\left(\begin{aligned} j_{|\kappa+1/2|-1/2}(i p r) 
\\ 
i \frac{pe^{-i\theta}}{M + E} \frac{\kappa}{|\kappa|} j_{|\kappa-1/2|-1/2}(i p r)
\end{aligned}\right),
\end{equation}
and
\begin{equation}
\begin{pmatrix} G^\infty_{E\kappa,\theta} \\ F^\infty_{E\kappa,\theta} \end{pmatrix} =
r e^{i\theta} 
\left(\begin{aligned} h^{(1)}_{|\kappa+1/2|-1/2}(i p r) 
\\ 
i \frac{pe^{-i\theta}}{1 + E} \frac{\kappa}{|\kappa|} h^{(1)}_{|\kappa-1/2|-1/2}(i p r)
\end{aligned}\right)
\end{equation}
with $p = \sqrt{e^{2i\theta}(1 - E^2)}$ and $Re(p) > 0$.
The Wronskian equals to
\begin{equation}
W = F^0_{E\kappa,\theta} G^\infty_{E\kappa,\theta} - G^0_{E\kappa,\theta} F^\infty_{E\kappa,\theta} 
= -\left[pe^{-i\theta}(1 + E)\right]^{-1}.
\end{equation}
%
% ========================
\section{Complex-Scaled external-potential Dirac Green function}
\label{z_green}
As in the case of the free Dirac Greed function, the external-potential Dirac Green function can be expressed as follows
\begin{equation}
G_{\theta}(E,\br_1,\br_2) = -\sum_{\kappa m}
\left[ 
\Phi_{E\kappa m, \theta}^{\infty}(\br_1) \Phi_{E\kappa m, \theta}^{0\dagger}(\br_2)\theta(r_1 - r_2)
+
\Phi_{E\kappa m, \theta}^{0}(\br_1) \Phi_{E\kappa m, \theta}^{\infty\dagger}(\br_2)\theta(r_2 - r_1)
\right],
\end{equation}
where $\Phi$ stands for the solutions of the complex-scaled Dirac equation in the external potential
\begin{equation}
\begin{pmatrix}
\frac{d}{dr} + \frac{\kappa}{r} && -e^{i \theta} \left[1 + E - V(re^{i\theta})\right]
\\
-e^{i \theta} \left[1 - E + V(re^{i\theta})\right] && \frac{d}{dr} - \frac{\kappa}{r}
\end{pmatrix}
\begin{pmatrix} G_{E\kappa,\theta} \\ F_{E\kappa,\theta} \end{pmatrix} = 0.
\end{equation}
To solve these equations we use the numerical approach described in Ref.~\cite{Yerokhin_PRA83_012507}.
The solution regular at the origin $\Phi^0_{E\kappa m,\theta}$ is constructed by direct propagation from $r=0$.
To obtain $\Phi^\infty_{E\kappa m,\theta}$ we propagate from $r > R$ backwards to $r = 0$.
Here $R$ is defined by the requirement that at $r > R$ the potential turns to the pure Coulomb one.
For that region, the explicit form of the regular at infinity solution is
\begin{equation}
\left(\begin{aligned}
G^\infty_{E\kappa,\theta} \\ F^\infty_{E\kappa,\theta}
\end{aligned}\right)
 =
\frac{1}{\sqrt{2pr}}
\left(\begin{aligned}
\left(\kappa +  \nu\frac{1}{E}\right) W_{\nu-1/2,\gamma}(2pr) + W_{\nu+1/2,\gamma}(2pr)
\\ 
\frac{pe^{-i\theta}}{1 + E}
\left[
\left(\kappa +  \nu\frac{1}{E}\right) W_{\nu-1/2,\gamma}(2pr) - W_{\nu+1/2,\gamma}(2pr)
\right]
\end{aligned}\right),
\end{equation}
where $p = \sqrt{e^{2i\theta}(1 - E^2)}$ with $Re(p) > 0$ and
\begin{equation}
\nu = \frac{\alpha Z E e^{i\theta}}{p}
\end{equation}
is the Sommerfeldt parameter.
By normalizing the obtained solutions to the unity Wronskian, one obtains the complex-scaled external-potential Green function.

%
% ======================================
%
\section{Gaussian nuclear charge distribution}
Here we utilize the Gaussian nuclear charge distribution
\begin{equation}
V_{\rm nuc}(re^{i\ctheta})
=
-\frac{\alpha Z}{re^{i\ctheta}}
\erf \left(\sqrt{\frac{3}{2}} \frac{re^{i\ctheta}}{R_{\rm nuc}} \right)
\end{equation}
with $Z$ standing for the nuclear charge and $R_{\rm nuc}$ is the nuclear charge radius.
The choice of this potential is provided by the fact that it can be dilated into the complex plane for a wide range of $\ctheta$.
We are aware that this potential represents a very rough approximation to the exact two-center one. 
The calculations beyond the monopole approximation are expected to provide only minor quantitative changes~\cite{Maltsev_PRA98_062709_2018}.
\subsection*{Zero-potential and one-potential  contributions}

The regularized zero- and one-potential terms are given by
\begin{equation}
\Delta E^{(0)}_a = 
\int \frac{d\bp}{(2\pi)^3} 
\bar{\psi}_a(\bp) \Sigma_R^{(0)}(\varepsilon_a;\bp) \psi_a(\bp)
\end{equation}
and
\begin{equation}
\Delta E^{(1)}_a = 
\int \frac{d\bp}{(2\pi)^3} \int \frac{d\bp'}{(2\pi)^3}  
\bar{\psi}_a(\bp')\,V(|\bp'-\bp|) \Gamma_R^{0}(\varepsilon_a;\bp',\bp) \psi_a(\bp),
\end{equation}
where the Fourier transform of the wave function $\psi_a$ coordinate representation is defined as
\begin{equation}
\psi_a(\bp) = \int d\br e^{-i\bp\cdot\br} \psi_a(\br).
\label{eq_fourier}
\end{equation}
The explicit expressions for $\Sigma^{(0)}_R$ and $\Gamma_R^{0}$ can be found in Refs.~\cite{Yerokhin_PRA60_800, Oreshkina_PRA101_032511_2020}.
Note that if the uniform complex rotation is applied, one needs also to modify Eq.~\eqref{eq_fourier} by transforming the radial coordinate in accordance with  the rule $\br \rightarrow \br \, e^{i\ctheta}$.
After this transformation, however, the integral will diverge exponentially.
To overcome this issue, we perform the uniform complex rotation in the momentum space on an opposite angle,
$p \rightarrow p e^{-i\ctheta}$.

The regularized expression for the zero-potential term, being integrated over the angular variables, is then given by 
\begin{align}\label{eq_zero_pot_cs}
\Delta E^{(0)}_a  &= 
e^{-3i\ctheta}
\frac{\alpha}{4\pi}
\int_0^\infty \frac{p^2dp}{(2\pi)^3}
%\left\lbrace
\biggl\{a(\rho_a)\left[\tg_a^2 -\tf^2_a\right] 
+
b(\rho_a)\left[\varepsilon_a\left(\tg_a^2 + \tf^2_a\right) + 2pe^{-i\ctheta}\tg_a\tf_a \right]%\right\rbrace,
\biggr\}
\end{align}
where 
\begin{equation}
\rho_a \equiv \rho_a(pe^{-i\theta}) = 1 - \varepsilon_a^2 + p^2e^{-2i\ctheta}
\end{equation}
and the functions $a(\rho)$ and $b(\rho)$ are defined as in Refs.~\cite{Yerokhin_PRA60_800, Oreshkina_PRA101_032511_2020}.
The momentum representation of the large $g_a$ and small $f_a$ components of the radial wave function are respectively defined as
\begin{align}
\tg_a & \equiv \tg_a(pe^{-i\theta}) 
= 
e^{3i\theta}\sqrt{\frac{2}{\pi}} \int_0^\infty dr r^2 j_l(pr) g_a(re^{i\theta}),
\\
\tf_a & \equiv \tf_a(pe^{-i\theta})
= 
-e^{3i\theta}\frac{\kappa}{|\kappa|}\sqrt{\frac{2}{\pi}} \int_0^\infty dr r^2 j_{\bar{l}}(pr) f_a(re^{i\theta}).
\end{align}
Here $\kappa = (-1)^{j+l+1/2}(j + 1/2)$ is the Dirac quantum number with $j$ and $l$ standing for the total and orbital angular momenta, respectively, $\bar{l} = 2j-l$, and $j_l$ is the spherical Bessel function.

\end{document}